Jared S. Ginsberg[1,*], Adam C. Overvig[1], M. Mehdi Jadidi[1], Stephanie C. Malek[1], Gauri N. Patwardhan[1,2], Nicolas Swenson[2], Nanfang Yu[1], and Alexander L. Gaeta[1]

# Enhanced Harmonic Generation in Gases Using an All-Dielectric Metasurface

**Abstract:** Strong field-confinement, long-lifetime resonances, and slow-light effects suggest that metasurfaces are a promising tool for nonlinear optical applications. These nanostructured devices have been utilized for relatively high efficiency solid-state high-harmonic generation platforms, four-wave mixing, and Raman scattering experiments, among others. Here we report the first all-dielectric metasurface to enhance harmonic generation from a surrounding gas, achieving as much as a factor of 45 increase in the overall yield for Argon atoms. When compared to metal nanostructures, dielectrics are more robust against damage for high power applications such as those using atomic gases. We employ dimerized high-contrast gratings fabricated in silicon-on-insulator that support bound states in the continuum, a resonance feature accessible in broken-symmetry planar devices. Our 1D gratings maintain large mode volumes, overcoming one of the more severe limitations of earlier device designs and greatly contributing to enhanced third- and fifth-harmonic generation. The interaction lengths that can be achieved are also significantly greater than the 10's of nm to which earlier solid-state designs were restricted. We perform finite-difference time-domain simulations to fully characterize the wavelength, linewidth, mode profile, and polarization dependence of the resonances. Our experiments confirm these predictions and are consistent with other nonlinear optical properties. The tunable wavelength dependence and quality-factor control we demonstrate in these devices make them an attractive tool for the next generation of high-harmonic sources, which are anticipated to be pumped at longer wavelengths and with lower peak power, higher repetition rate lasers.

## 1 Introduction

Strong laser field excitation of rare gas atoms has been the preferred method of generating short wavelength coherent radiation down into the soft x-ray range [1–5]. High-order harmonic generation (HHG) has been a pillar of ultrashort pulse generation for decades [6,7], and has been the catalyst for significant advances in nonlinear optics and attosecond science. Next-generation studies of ultrafast atomic phenomena demand higher repetition rates and shorter pulses with greater bandwidths. Thus, the quadratic dependence of the maximum HHG bandwidth on pump wavelength [8] provides strong motivation to push the wavelength of the pump field deeper into the mid-infrared. Strict intensity requirements of nearly 100 terawatt/cm$^2$ in order to access the non-perturbative regime of HHG in gases has, however, mostly restricted the field to the use of chirped pulse, multi-pass, and regeneratively amplified lasers. These large-footprint systems have low repetition rates, high average power, and are limited to the near-visible regime, hindering progress towards mid-infrared pumping. Our work is motivated by a simultaneous pursuit of efficient HHG at longer wavelengths and lower pump powers.

The lower intensities required to observe HHG from solids [9] has generated significant interest and led to rapid innovation in the decade since its initial discovery. A promising direction has been the merging of nonlinear optics with metasurfaces to create an efficient platform for HHG [10–15]. These nanostructured devices allow for the engineering of far-field emission profiles, selective wavelength enhancement, and exceptionally strong field-confinement. Sub-wavelength scale plasmonics and dielectrics have both been demonstrated for these purposes. While the strong field-confinement permitted by these two types of systems further reduces the pump power requirements for harmonic generation in solid-state systems, enhancing intensities within the device material poses its own challenges. Plasmonic devices are severely hampered by ohmic losses and are susceptible to melting, even at modest intensities [16]. Both plasmonics and dielectrics must also contend with the reabsorption of harmonics within an opaque generation material [17], reducing the effective interaction length of the nonlinear medium to only a few tens of nanometers. Alternatively, spatially selective confinement of pump energy within the regions just outside the metasurface can be accomplished with bound states in the continuum (BIC), a class of optical resonance with broadly engineerable mode profiles supported by all-dielectric metasurfaces [18–21]. A variety of BIC-enhanced light-matter interactions including optical absorption [22], third-harmonic generation (THG) in solids [23], and Raman scattering [24] have been reported.

In this work we demonstrate an all-dielectric metasurface to significantly enhance the harmonic yield from gas-filled gaps by utilizing dimerized high-contrast gratings (DHCGs) supporting strong BIC resonances near 1550 or 2900 nm. DHCG metasurfaces offer independent parameters for band structure design, wavelength selection, and quality factor tuning [25], making them well suited for HHG. We perform experiments that confirm the mode location, polarization dependence, and resonance wavelength predicted by our device simulations and show that harmonic emission from Argon atoms can be enhanced by more than an order of magnitude via resonant excitation of our devices.

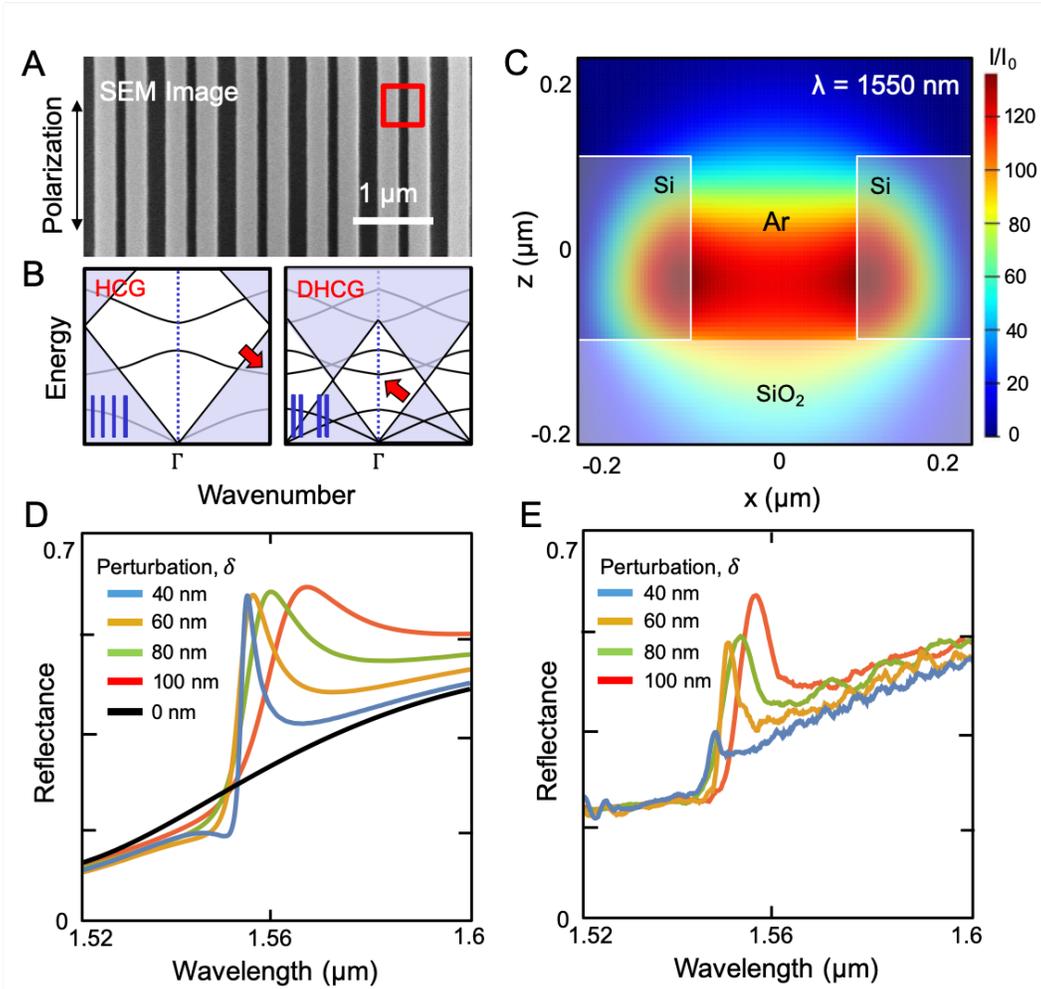

**Figure 1:** Device fabrication and linear characterization.
(A) SEM image of the fabricated dimerized high-contrast grating (DHCG). The height of the grating is 250 nm, the finger width is 270 nm, and the period is 920 nm. The dimerizing perturbation to the grating finger spacing results in alternating smaller gaps of 150 nm and larger gaps of 230 nm. (B) Typical band structures of unperturbed HCG (left) and perturbed DHCG (right). Red arrows indicate the mode shown in (C) which is brought to the gamma point by Brillouin zone folding. Shaded blue regions are below the light-line. (C) FDTD simulation showing the relative power enhancement in the region marked by a red square in (A), now a side view. The strong field-confinement mode is centered in the Argon gas region, with leakage into the walls of the grating fingers and substrate, outlined in white. (D) Simulated unpolarized reflectance spectrum of DHCGs with alternating gaps of 190 nm $\pm\ \delta$. The reflectance peaks at the BIC Fano-resonance design wavelength near 1550 nm. Smaller perturbations correspond to larger Q-factors and narrower linewidths. (E) FTIR measured reflectance of unpolarized near-infrared light. Losses lead to the observed variation in reflectance magnitude.

## 2 Device Principles

Bound states in the continuum were first proposed as a mathematical curiosity of quantum mechanics [26], but have since been realized in a number of experimental environments beyond quantum systems. In periodic, planar optical devices BICs manifest as infinitely strong resonances of zero linewidth [27], and are therefore not experimentally accessible. One such example is a normally incident plane-wave or Gaussian beam and a high-contrast grating with an antisymmetric mode at its gamma point, that is, wavevector $k = 0$. The inability of an even symmetry incident wave to couple to an odd symmetry device mode is the working principle of a so-called symmetry-protected BIC [28]. Coupling to such a resonance therefore requires either the breaking of the even symmetry of the incident beam or of the odd symmetry of the mode, a process that results in a state that is quasi-bound in the continuum (a quasi-BIC). The former requires non-normal incidence of the pump beam, while the latter can be achieved by a periodic perturbation to the grating unit cell. The type and magnitude of the perturbation offer significant flexibility when selecting mode profiles and device bandwidth [29].

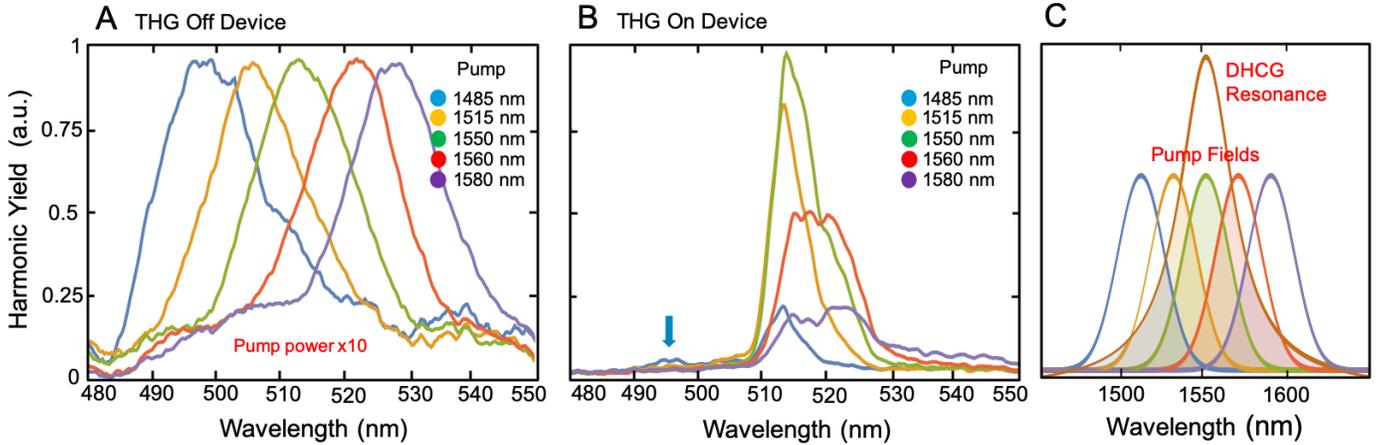

**Figure 2:** Near-infrared device wavelength characterization.
(A) Normalized third-harmonic signal generated from the bare substrate for pump wavelengths in the range of 1485 nm to 1580 nm. Pump powers are 10X greater than in B. (B) Third-harmonic spectra for the same five pump wavelengths shown in (A), now on the DHCG. The large field enhancement on resonance pins the third harmonic to 513 nm. The small signal at 495 nm (blue arrow) corresponds to the third-harmonic signal from the peak of the 1485-nm pump that is generated off-resonance. (C) Schematic of the pump field and resonance overlap resulting in the harmonics measured in B. The shaded overlap regions contribute to the nonlinear signal.

We designed and fabricated silicon-on-insulator (SOI) DHCGs on a standard wafer with a 250-nm thick silicon device layer, 1-μm of oxide, and a 500-μm silicon substrate. Full-wave device simulations were performed with Lumerical, a commercial finite-difference time-domain (FDTD) software package. Figure 1A shows a top view SEM image of the finalized device, fabricated by e-beam lithography, which consists of a series of silicon fingers 270-nm wide with a total grating period of 920-nm. The large grating period is a result of the dimerization of an unperturbed high contrast grating with a period of 460-nm. We also fabricated a second grating with approximately twice the width and spacing, allowing us to test two very different operating wavelengths of 1550 nm and 2900 nm. By scaling the period and duty cycle, the system can be made to operate at any wavelength and is limited only by fabrication sensitivity.

The dimerization of a high-contrast grating alters the physics of the system in ways beyond the perturbation to the mode symmetry described above. In the momentum space picture, the doubling of the spatial period of the structure implies a halving in k-space, known as Brillouin zone folding [30]. This "folds" modes at the edges of the first Brillouin zone (FBZ) to the gamma point, making them accessible to normal incidence excitation. Such modes had previously only existed in an unusable state below the light-line. The Brillouin zone folding of a high contrast grating is illustrated in Figure 1B. In a 1D DHCG, such as the ones described in this work, two types of periodic perturbations can be implemented to dimerize the unit cell. The finger width or spacing is alternatively increased and decreased by a perturbation $\delta$, while the other parameter is held constant. We choose to modulate the spacing, (i.e., the gap-perturbation), since this folds the unique mode depicted in Figure 1C to the gamma point. The mode is centered in the gap between adjacent grating fingers, where it provides a uniform two orders of magnitude enhancement in the local field intensity for a modest quality factor Q = 1500. This mode affords us the ability to inject a gaseous nonlinear medium (e.g., Argon atoms) into the gaps. Only when the polarization of the pump is oriented along the finger direction can this particular TE mode be efficiently excited. The total number of modes in the system is also a function of the grating depth. By restricting the device to 250 nm in height we thereby reduce the likelihood of a parasitic resonance being located near our gap-centered mode. Alternatively, a mode centered purely within the silicon could have been achieved with a finger-width perturbation and fixed gaps.

The modes that exist at the Brillouin zone boundaries are typically much flatter than modes found at the gamma point, and the slope of the band is directly proportional to the group velocity of light in the mode [9,10]. Therefore, high degrees of confinement in small devices benefit from the flattest possible bands. Furthermore, the combination of small slopes and a reduced FBZ means DHCGs can exhibit sharper spectral features with narrower linewidths than either unperturbed or low-contrast gratings. However, for all DHCG modes the quality factor of the resonance can be fully tuned by the magnitude of the dimerizing perturbation according to $Q \propto \delta^{-2}$ for small perturbations $\delta$. The inverse quadratic scaling of the Q-factor is a general feature of all quasi-BIC supporting metasurfaces [33]. Figures 1D and 1E show the simulated and measured unpolarized reflectance spectra of a set of DHCGs designed with a central wavelength of 1550 nm and a range of Q-factors as high as 1000. We also include the symmetry-protected case of unperturbed gap widths of 190 nm and confirm that no resonance is observed. In our experiments we choose to increase the size of the perturbation in order to accommodate the bandwidth of our laser pulses, while simultaneously making it easier for the field to couple into the mode.

# 3 Experiments

Our experimental setup utilizes an optical parametric amplifier (Light Conversion HE Topas Prime) pumped by an amplified Titanium-sapphire laser system (Coherent Legend Elite) operating at a 1-kHz repetition rate with 6 mJ of pulse energy to generate tunable, 60-fs duration signal pulses with center wavelengths from 1485 nm to 1580 nm. This parametric amplifier output is used to pump an additional difference frequency generation module for our mid-infrared measurements at 2.9 μm with a pulse duration of 70 fs. The pump-beam polarization is controlled with a broadband zero-order half wave plate before being focused on the sample by a 5-cm focal length $CaF_2$ lens. Any unconverted seed light is removed with the appropriate long-pass filters. Reflected harmonics are collected and measured on a fiber-coupled ultraviolet-visible spectrometer (Ocean Optics) with a 350 nm to 1100 nm detection range. The devices are mounted in a high-pressure stainless-steel gas cell with a 1-inch diameter sapphire window. For sufficiently large beam diameters, the intensity on the sapphire window is low enough that it does not contribute to the nonlinear signal in our experiments. Argon pressures in excess of 200 psi could be held on the devices within the cell.

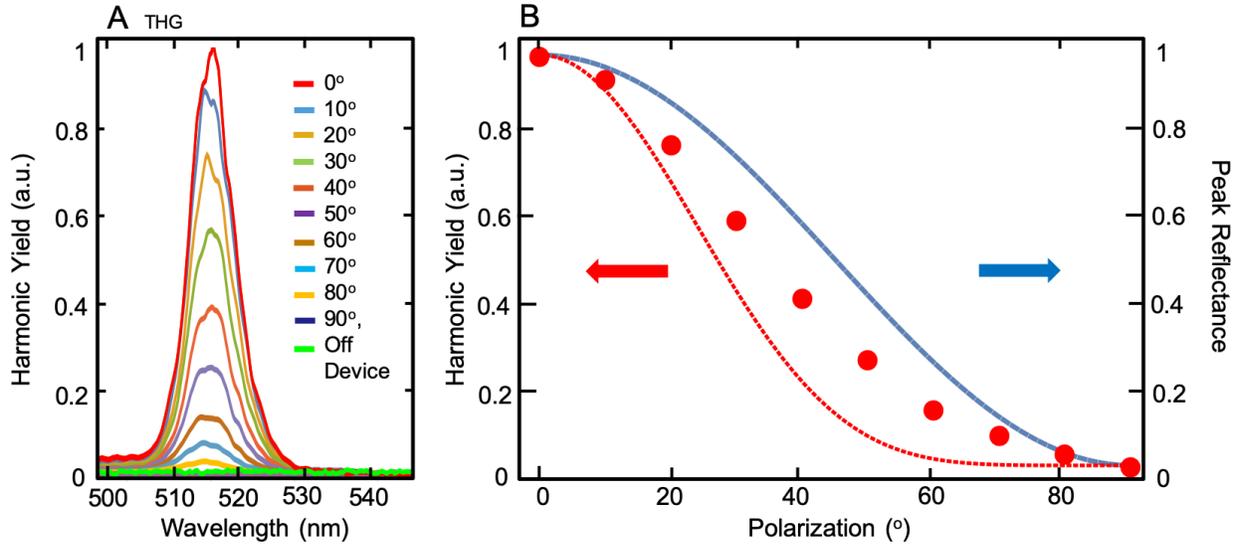

**Figure 3:** Grating polarization characterization.
(A) Normalized third-harmonic spectra of a 1550 nm pump, corresponding to the red points in B. Also included is the signal generated off the device, which is unmeasurable. 1 atmosphere of Argon pressure was flowed over the sample. Contributions to the third harmonic from both gas and silicon are present. (B) Simulated peak reflectance of the ideal grating (blue line), and integrated experimental third harmonic yield (red dots), as a function of the pump polarization. 0° is the polarization parallel to the grating fingers. The red dashed line indicates $\cos^6\theta$, corresponding to the expected scaling of perturbative THG. The mismatch between the red theoretical curve and experimental dots indicates that the measured THG is instead following a non-perturbative scaling typical of higher pump intensities.

The third-harmonic-generation (THG) signals generated on and off the 1550 nm- resonant DHCG are shown in Figure 2. Off the metasurface structure, the pump wavelength was swept from 1485 nm to 1580 nm, and the emitted third harmonic signal was recorded in Figure 2A. We observe the expected behavior of harmonic emission that tracks the pump wavelength without any additional spectral features. This picture changes when we repeat the same wavelength scan on the device itself. The harmonic yield dramatically increases, requiring over an order of magnitude lower power to generate, but within a limited bandwidth. At pump wavelengths below 1530 nm or above 1580 nm no measurable increase is observed in the third-harmonic signals plotted in Figure 2B, and only a small signal, far off-resonance due to the peak of the 1485 nm pump is measured (blue arrow). Compared to this small signal, the 1550-nm pump beam displays the greatest harmonic yield enhancement of a factor of 45, implying the greatest overlap with the BIC resonance. This concept is illustrated in Figure 2C, in which the overlap of the Gaussian pump spectra and Lorentzian BIC resonance determines efficient harmonic generation. This explains the observed pinning of the THG to roughly 516 nm and the reduced linewidths of detuned harmonics. The cut-off shape of the short wavelength harmonics gives the clearest indication of the resonance edge.

We also examine the polarization dependence of our fabricated near-IR metasurfaces. As described in the previous section, the grating mode under study is excited by TE polarized light parallel to the long direction of the grating fingers. We hereafter refer to this polarization direction as 0°. The theoretical dependence of the peak reflectance from the grating is studied in a series of FDTD simulations and plotted as a blue line in Figure 3B. For clarity, we subtract any contribution to the reflectance due to etalon effects of the thin device and oxide layers, hence the reflectance has a minimum of zero. As the polarization is rotated by an angle $\theta$ beginning from 0°, the peak power reflectance due to the BIC follows precisely the expected $\cos^2\theta$ behavior. The component of the field oriented along 90° therefore makes no contribution, and the $\cos^2\theta$ behavior is a result of the projection of the pump beam onto the 0° axis. We compare to this reflectance curve the integrated (measured) THG yield of 1550-nm pulses on resonance, shown as red dots in Figure 3B, corresponding to the spectra in Figure 3A. The experimental data follow a similar trend, but with some important differences. Were it a perturbative third-order nonlinearity with $I \propto \cos^2\theta$, a THG efficiency proportional to $\cos^6\theta$ may be expected, given by the red line in Figure 3B. The measured harmonics follow neither the exact $\cos^2\theta$ behavior of the resonance or the perturbative $\cos^6\theta$ dependence. The observed saturation of the intensity dependence of the $N^{th}$ order harmonic below $I^N$ is a signature of non-perturbative harmonic generation. Thus, we conclude that in these experiments the generated signal corresponds to the non-perturbative regime with more than 32 times larger emission at 0° than at 80°. Within the nonperturbative regime for THG pumped in the mid-IR, a conversion efficiency upper-bound of about $10^{-7}$ is found for low-pressure Argon gas [34]. Fifth harmonic conversion efficiencies saturate about one order of magnitude lower at $10^{-8}$. While the intensity enhancement provided by the DHCGs demonstrated here does not raise this efficiency limit, it does successfully lower the input power at which the maximum conversion efficiency is reached. At 90° polarization, the harmonic yield converges back to the signal off-resonance.

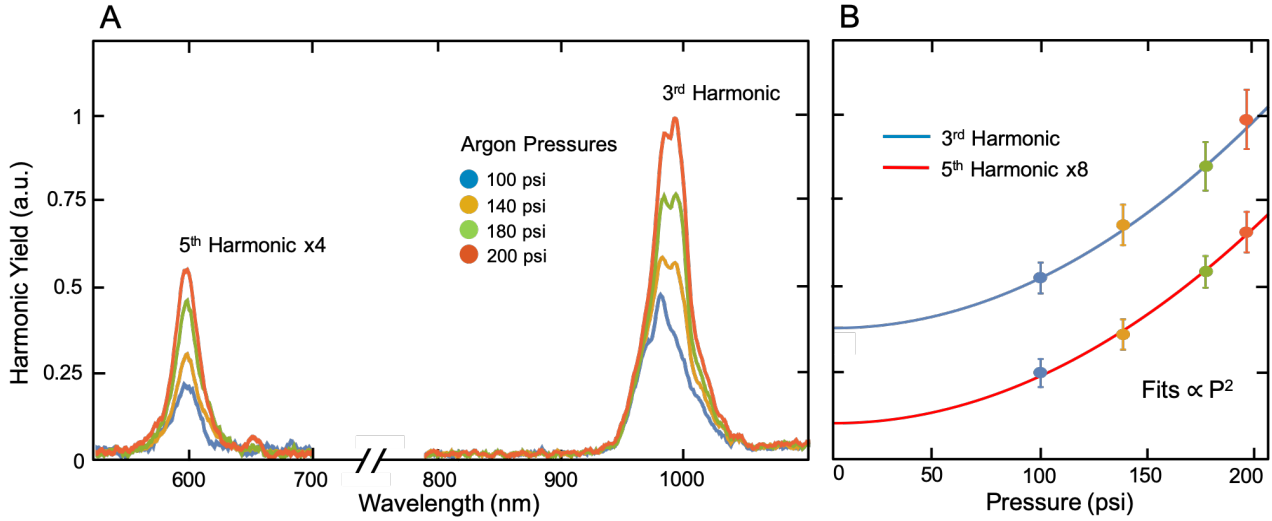

**Figure 4:** Pressure dependence of mid-infrared device.
(A) Third and fifth harmonics of a 2.9 μm pump beam on a 2.9 μm resonant DHCG. Argon pressures are varied from 100 psi to 200 psi in a high-pressure gas cell. The fifth harmonic has been scaled up for visibility. (B) Integrated harmonic yields from the spectra in (A), matched by color. Pressure dependence of both harmonic orders are fit to the square of the pressure added to a pressure independent contribution that we assume is due to the silicon fingers. The fifth harmonic is similarly scaled up.

We also characterize the performance using the scaled-up mid-infrared devices, allowing us to explore the third and fifth harmonics, both of which fall within our detection range for a 2.9 μm pump wavelength. Figure 4 shows the third- and fifth-harmonic spectra of these structures for a range of Argon gas pressures. For both harmonic orders, the harmonic yields scale quadratically with gas pressure (see Fig. 4B) confirming that the emission is due to the atoms in the gap regions. At a pump intensity of $2 \times 10^{12}$ W/cm$^2$, this emission occurs at field strengths where efficient harmonic generation in Argon is not expected to take place. In the absence of Argon, we still observe a non-zero third- and fifth-harmonic signal, which we attribute to the silicon finger surfaces. This is further supported by the expected overlap of the mode and grating predicted in our simulation (Figure 1C). At 200 psi of Argon pressure, the harmonic yields from Argon significantly exceed that of the device material.

## 4 Discussion and Outlook

We have observed a greater than 10-fold enhancement of the third and fifth harmonic yields from Argon atoms in the presence of a BIC- supporting DHCG, with the yield enhancement for higher-order harmonics being greater due to the higher order dependence on pump intensity. We have evidence based on the polarization dependence of THG depicted in Figure 3 that we have achieved the strong-field, non-perturbative regime, as the integrated THG does not follow a $\cos^6\theta$ dependence on the polarization angle. From this observation we note that the hybrid gas-DHCG technique lowers the required pump input intensity to saturate the THG conversion efficiency at about $10^{-7}$. Furthermore, we have determined that for pressures between 100 psi and 200 psi, the contributions to our total signal from enhanced harmonics within the grating material and Argon are of the same order of magnitude. With this we confirm the mode profile produced by our FDTD simulations which predicts a gap-centered mode with leakage into the silicon fingers.

Further improvement of our hybrid metasurface-gas harmonic scheme requires the ability to restrict the strong field-confinement mode to only the gas-filled region. Ideally, this would be accomplished while maintaining relatively large mode areas and all of the tunable wavelength and quality factor properties afforded to us by the DHCG platform. We propose a device in Figure 5, based on a 2-dimensional realization of a DHCG, that meets all of these stated requirements. Figure 5A shows a representation of the device, which consists of silicon dimers containing one circular and one elliptical nanopillar. The perturbation of the unit cell is to the ellipse minor axis. For a target resonance wavelength in the mid-infrared, the disk major axes should be 500 nm, and the grating period should be 2.1 μm. Figure 5B shows the BIC mode of interest in the system, chosen at a wavelength of 3.15 μm. In this system, the strongest field enhancement no longer spatially overlaps with the silicon. This offers tremendous flexibility to the proposed system in two forms. First, the contribution to the harmonic signal from any non-gas sources would be greatly diminished. Second, the quality-factors could be raised to their fabrication limits, greatly increasing the local powers experienced by the atomic targets without increasing the likelihood of device damage. Quality-factor tuning of the 2D DHCG observes the required $\delta^{-2}$ dependence, but now the lower-Q peaks in the reflectance curve tend toward the short wavelength side as in Figure 5C. Smaller ellipse minor axes corresponding to larger perturbations result in less silicon in each unit cell, and hence a lower effective index, which leads to a blueshift in the reflectance peak.

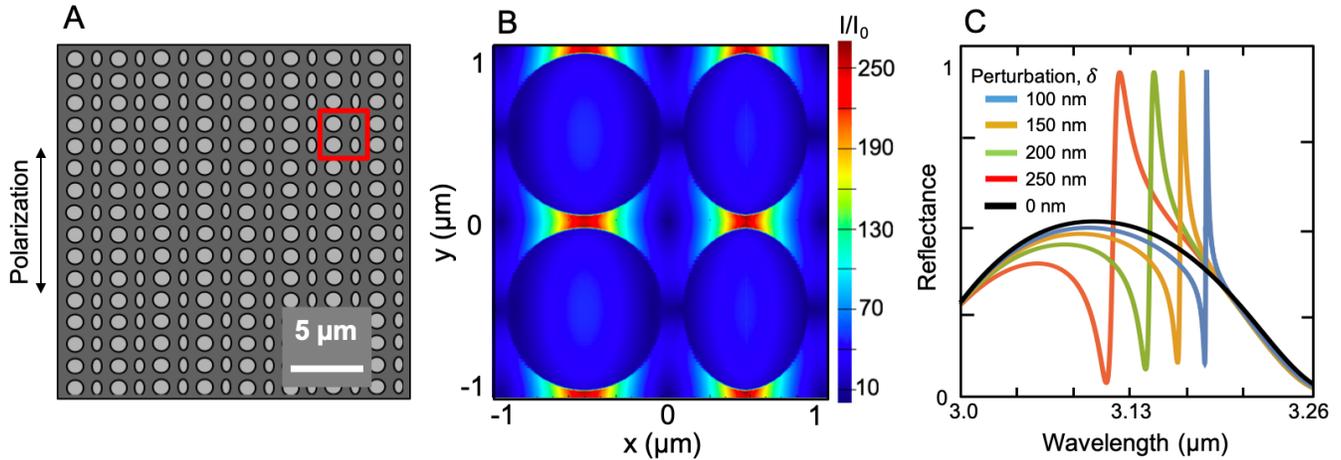

**Figure 5:** 2D DHCG Simulation
(A) Proposed 2D DHCG. The height of the grating is 250 nm, disk radii 500 nm, and the period is 2.1 μm. The dimerizing perturbation is to every other disk's minor axis. (B) FDTD simulation showing the relative power enhancement in the region marked by a red square in (A) (top view). The strong field-confinement mode is entirely contained in the Argon gas region, with no leakage into the walls of the grating or substrate. (C) Simulated polarized reflectance spectrum of 2D DHCGs with alternating minor axes of 500 nm $-\delta$ for four magnitudes of the perturbation as well as the unperturbed case. The reflectance peaks at the BIC Fano-resonance wavelength near 3.15 μm. Smaller perturbations correspond to larger Q-factors and narrower linewidths.

The potential for greatly amplified field strengths within gas-filled gaps gives renewed relevance to high-harmonic generation from gases in the presence of metasurfaces. Since a critical factor in the generation of high energy harmonics is, among other things, a sufficiently large intensity, the approach we have developed in this work can be extended readily to higher-order harmonic generation. The possibility to tailor mode profiles to exclude leakage into the grating, and therefore eliminate the damage that has so far limited plasmonic and dielectric devices alike, deserves further study. Using a transparent nonlinear medium such as Argon extends the interaction length of the harmonic generation process well beyond the 10's of nanometers available in either metal or silicon-based sources, while dramatically reducing harmonic absorption when compared with solid-state media. The 250-nm tall gratings used in our experiments, for example, are tall enough to allow greater coherent buildup of the harmonic signal, while being sufficiently subwavelength in extent to sidestep phasematching restrictions. Lower power requirements open up the possibility to use higher repetition rate sources to initiate HHG in gases and have the potential to make smaller footprint table-top attosecond sources more viable. The ease with which Q-factors can be adjusted in the fabrication process means the DHCG devices shown here can be made to match the bandwidth of a given pump laser. Alternatively, the resonance linewidth can be chosen to select for the desired linewidth of the generated harmonics. Moreover, the control provided by DHCGs can be further extended to simultaneously include both wavefront and polarization shaping.

Looking to the future, our hybrid gas-DHCG scheme could be applied to the entire suite of nonlinear optical studies being conducted in atomic systems. Beyond the gas phase, two-dimensional materials are emerging as a promising nonlinear optical platform moving forward [35–37]. The single- to few- atom thickness of graphene and transition metal dichalcogenides limits their nonlinear interaction length, making them prime candidates for integration with field-confining metasurfaces. Furthermore, the all-dielectric metasurface is a leading platform for meta-lenses, capable of providing the same level of phase-control and integrability with 2D materials as seen in previous plasmonic nonlinear metalenses [38]. This is due to the high degree of engineerability of modes provided by all-dielectric DHCGs, but they also bring the added benefit of eliminating potential field hotspots in sensitive 2D materials, especially at higher harmonic orders and greater field intensities.